hep-lat/9501001 2 Jan 1995

# Dependence of Lattice Hadron Masses on External Magnetic Fields

H.R. Rubinstein ◇, S. Solomon⋆ and T. Wittlich•

◇ *Department of Theoretical Physics, Uppsala University, S-75121 Uppsala, Sweden, e-mail: rub@tsl.uu.se*

⋆ *Racah Institute of Physics, Hebrew University of Jerusalem, Jerusalem, Israel, e-mail: sorin@vms.huji.ac.il*

• *Physikalisches Institut, Universität Bonn Nußallee 12, D-53115 Bonn, Germany e-mail: thomas@theoa1.physik.uni-bonn.de*

**Abstract**

We study the variation of the hadron masses in the presence of external magnetic fields of strength of the order of the masses themselves.

We identify the main factors affecting the lattice simulation results:

- the boundary discontinuities for $eB << 2\pi/L^2a^2$.
- the SU(6) choice of the hadron wave-function.

We confirm qualitatively the earlier theoretical ansatz on the linear behaviour of the masses with the magnetic field and, as a by-product, we improve the lattice measurements of the nucleon magnetic moments.

However our systematic and statistical errors preclude us from measuring the theoretically predicted field strength at which the proton becomes heavier than the neutron.

---

# 1 Introduction

We study the variation of the masses of the proton and the neutron in the presence of a uniform external magnetic field using lattice QCD.

The behavior of proton, neutron and electron in the presence of a constant magnetic field $B$ was considered in [1].

For a sufficiently small magnetic field, it was assumed in [1] that the dependence of the hadron masses on $B$ is controlled by their anomalous magnetic moments which are known from experiment. The masses of proton and neutron can be easily computed using Landau's formula for a Dirac particle in a constant magnetic field:

$$E_s^p = m_p + \frac{|eB|}{2m_p} - g_p \frac{eB}{2m_p} s \tag{1.1}$$

$$E_s^n = m_n - \frac{eB}{2m_n} g_n s \tag{1.2}$$

$s \in \{-1, +1\}$ denotes the spin, $m_p$ and $m_n$ the masses in the absence of $B$. The formulae hold for zero momentum and the lowest Landau level.

Note that the $|eB|$ coefficients ($1/2m_p$ and respectively 0) were considered non-anomalous.

If one inserts the experimentally known anomalous magnetic moments, $g_n = -1.9$ and $g_p = 2.79$, then for increasing magnetic field the lowest state of the neutron goes down steeper than the lowest state of the proton. Consequently since in nature the neutron is heavier than the proton there exists a value of the magnetic field $B_0$ where the masses of the two particle are equal. $B_0$ is a very large field of the order of $10^{14}T \approx 10^{-2}GeV^2$. For magnetic fields larger than $B_\beta \approx 10^{16}T$ the proton, neutron and positron acquire masses which allow the $\beta$–decay of the proton.

Magnetic fields of that order of magnitude may appear in certain astronomical phenomena [2, 3] and an unstable proton would be a very interesting phenomenon.

However the magnetic field required is very large so that the assumption of pointlike particles does not necessarily hold and the effect of the strong forces has to be taken into account.

Since there is no analytic nonperturbative approach in QCD it is worthwhile to treat the problem numerically by means of lattice QCD.

External magnetic fields have already been used as a tool to compute g–factors of proton and neutron [4, 5]. In the present paper we study the spin up/down states of proton and neutron and compare them which the phenomenological approach (eq. (1.1) and (1.2)). In particular we compute the slopes (Energy/$B$) for the magnetic fields of interest. This is harder then the computation of the g–factors where it is sufficient to measure the mass *difference* of spin up and down states which is more stable against systematic and statistical errors.

We use a method similar to the one described in [4, 5]: The magnetic field in the $z$–direction is applied to a set of $SU(3)$ configurations by multiplying (2.1) the links by appropriate $U(1)$ phases (eq. (2.6) or 2.8)). In this way we define the gauge background for computing the propagator (2.10) of a quark with charge $e_q$. These propagators are then used in order to compute the correlation functions for proton and neutron (2.12).

The exponential time dependence of these correlators define the nucleon masses $m$ in the magnetic field:

$$m = -\lim_{t\to\infty} \frac{1}{t} \log \frac{G(t)}{G(0)} \qquad (1.3)$$

We find a dependence of the results on the different possible choices of the magnetic field configuration $U^B$ ((2.6), (2.8)). We describe how this problem is related to the gauge invariant definition of the charged hadron correlation functions (2.14). In order to get a definite procedure we consider first the 'flat' case where no QCD is present and tune the relevant parameters such that in the 'flat' limit our procedure reproduces the continuum theoretical results correctly (Table I). In order to estimate the reliability of our results, and their sensitivity to the wave function choice we compare the $SU(6)$ wave function for the neutron and proton correlators (2.12) with the $(u\gamma_5 Cu)d$ function [6] (Table II). While in the usual measurements there is no difference between the two definitions, in our context, the electric charges quantum numbers are as relevant as the color and spin ones in characterizing the wave function and significant differences are expected. This is discussed in detail in section 2. Section 3 presents the results: g–factors and slopes for proton and neutron (Table III, Figure 3).

We work with two lattice sizes and find a dependency of the results on the volume $V$. A linear extrapolation in $1/V$ (compare also [7]) leads to agreement with the phenomenological predictions (1.1) and (1.2).

The precision which was achievable does not allow a definite answer to the question whether the proton becomes heavier than the neutron in strong magnetic fields. However this work maps out the elements which play a role in the lattice simulation of hadrons in magnetic fields and opens the way for more massive studies.

In particular, our results are consistent with the main assumptions contained in the formulae (1.1) and (1.2):

1. The presence of QCD field has a drastic effect on the ground state vibration energy lowering the slope of $(m_{p\uparrow}+m_{p\downarrow})$ from $(2/3+2/3+1/3)|B|/m_q$ to $|B|/m_p$ and the slope of $(m_{n\uparrow} + m_{n\downarrow})$ from $(2/3 + 1/3 + 1/3)|B|/m_q$ to 0.

2. The QCD field has only a small effect on the splitting $(m_{p\uparrow} - m_{p\downarrow})/m_p$ leading to the small corrections to the anomalous magnetic moments of the independent quark model.

3. Note that we have nothing to say about the QCD mass difference $m_p(0) - m_n(0)$ in the absence of magnetic fields. This quantity is clearly totally beyond today's technology and we treat it as an external datum as the phenomenological treatment of [1] does.

More detailed conclusions are given in section 4.

# 2 The method

In this section we describe the ingredients of the method and how to tune them properly.

We work with a set of 30 quenched $SU(3)$ configurations generated at $\beta = 6.0$ by a variant of the Cabbibo–Marinari algorithm [8]. The configurations are separated by 500 sweeps with 2000 sweeps thermalisation. Statistical independence has been checked by comparing the variance of various Wilson loops for different bins of configurations. We use lattice sizes of $16 \times 16 \times 8 \times 8$ and $24 \times 16 \times 16 \times 12$ $(t \times x \times y \times z)$.

We work with Wilson fermions at $K = 0.1475$ and $0.1525$ for the small and $K = 0.145$ and $0.15$ for the large lattice. The quark propagators have been computed by means of a preconditioned conjugate gradient algorithm [9]. Since this has to been done for various values for $B$ and $e_q$ much more computer resources are needed compared to 'nonmagnetic' mass computations. We work with pointlike sources.

The remainder of this section is organized as follows:
Section 2.1 describes different realizations of the magnetic field and their implications on the results. Section 2.2 deals with the choice for the wave function appearing in the hadron propagator.

## 2.1 The magnetic field

On the lattice, a uniform magnetic field $B$ can be introduced by multiplying links $U_\mu(\tilde{x})$ with an external abelian phase $U_\mu^B(\tilde{x})$:

$$U_\mu(\tilde{x}) \rightarrow U_\mu(\tilde{x}) U_\mu^B(\tilde{x}) \tag{2.1}$$

where $\tilde{x} = (t, x, y, z)$.
In order to get a uniform magnetic field in z–direction one can set (ignoring for the moment the boundary conditions):

$$\begin{aligned} U_x^B(\tilde{x}) &= \exp(+ie\gamma y) \\ U_y^B(\tilde{x}) &= \exp(-ie\xi x) \\ U_z^B(\tilde{x}) &= 1 \\ U_t^B(\tilde{x}) &= 1 \end{aligned} \tag{2.2}$$

In that way the flux through a plaquette in the x–y plane is given by

$$\exp(ieBa^2) = \exp(i(\gamma + \xi)) \tag{2.3}$$

which defines the magnetic field as

$$a^2eB = \gamma + \xi \tag{2.4}$$

On a finite lattice with periodic boundary conditions (at least in direction x and y) , in order to get an uniform magnetic field, $B$ has to be a multiple of an integer $n$:

$$a^2eB = \frac{2\pi}{L_xL_y}n \tag{2.5}$$

To see that consider a possible realization for the magnetic field:

$$\begin{aligned} U_x^B(\tilde{x}) &= \exp(-ieBa^2yL_x) && \text{for } x = L_x - 1 \\ &= 1 && \text{else} \\ U_y^B(\tilde{x}) &= \exp(ieBa^2(x - x_0)) \\ U_z^B(\tilde{x}) &= 1 \\ U_t^B(\tilde{x}) &= 1 \end{aligned} \tag{2.6}$$

where $x = 0 \ldots L_x - 1$ etc.
$x_0$ is parametrizing the Polyakov loop degree of freedom.

According to (2.6) the plaquettes in the x–y plane are:

$$\begin{aligned} P^B(\tilde{x}) &= \exp(ieBa^2(-L_xL_y + 1)) && \text{for } x = L_x - 1, y = L_y - 1 \\ &= \exp(ieBa^2) && \text{else} \end{aligned} \tag{2.7}$$

Thus the magnetic flux is homogenous on all the x–y plaquettes with the possible exception of the corner $L_x - 1, L_y - 1$. In order that the corner plaquette will equal all the others, $B$ has to be quantized according to (2.5).

Unfortunately the lowest magnetic field fulfilling (2.5) is very large for reasonable sizes of the lattice: If we take for example $L_x = L_y = 8$ and $a^{-1} \approx 2GeV$ ($\beta = 6.0$) we get $eB \approx GeV^2$ (note that in practice the lowest field corresponds to a down quark with electric charge $e_d = -\frac{1}{3}e$). For a proton this corresponds to a mass shift $\delta m = \mu B$ of the order of one $GeV$ ($\mu$ = magnetic moment of the proton).
Thus in order to apply magnetic fields of interest we can either

- increase $L_x$ and $L_y$ or

- give up the quantization of $B$, which means that the field is not homogenous anymore.

Since increasing the lattice requires an amount of computer resources not available to us we have implemented two realizations of $B$ which give up quantization.

The first realization stems from the observation that only *one* plaquette of configuration (2.6) destroys the homogeneity if $B$ is not quantized. If this 'exceptional' plaquette is far from the source we may assume that its influence is not too strong. For that reason we place the source in the middle of the lattice $x_s = L_x/2$, $y_s = L_y/2$ (while the 'exceptional' plaquette resides at $x = L_x - 1$, $y = L_y - 1$) and increase the lattice size in the x–y plane. We refer to this realization as '1-ep' type ('one exceptional plaquette').

The second realization has been used in reference [4]. It applies *fixed* boundary conditions in the x–direction:

$$\begin{aligned} U_x^B(t,x,y,z) &= 0 && \text{for } x = L_x - 1 \\ &= 1 && \text{else} \\ U_y^B(t,x,y,z) &= \exp(ieBa^2(x-x_0)) \end{aligned} \tag{2.8}$$

The plaquettes in the x–y–plane are given by

$$\begin{aligned} P^B(t,x,y,z) &= 0 && \text{for } x = L_x - 1 \\ &= \exp(ieBa^2) && \text{else} \end{aligned} \tag{2.9}$$

We refer to this realization as 'fixed' type.

The freedom of choosing various Polyakov loop values was not exploited in [4]. However we will see that the choice of $x_0$ is significant in order to estimate the systematic errors originating in the finite size of the lattice and especially in the large fields related with the breaking of the quantization condition.

In order to estimate the systematic errors related to the lattice discretization, magnetic background representation and the wave function choices, we compare in the 'flat' case (all $SU(3)$ links set to one) our simulation results with predictions from the continuum.

We define the quark propagator

$$G_q(\vec{x},t;\vec{x_s},t_s) = <\bar{q}(\vec{x},t)q(\vec{x_s},t_s)> \tag{2.10}$$

(q=u,d for up and down quark) and the $SU(6)$ wave function [10] (e.g. for a proton with spin up)

$$
\begin{aligned}
\Psi_{p\uparrow}^{SU(6)} = \frac{1}{\sqrt{18}}( \quad & 2u_\uparrow(1)u_\uparrow(2)d_\downarrow(3) \\
+ \; & 2u_\uparrow(3)u_\uparrow(1)d_\downarrow(2) \\
+ \; & 2u_\uparrow(2)u_\uparrow(3)d_\downarrow(1) \\
- \; & u_\uparrow(1)u_\downarrow(2)d_\uparrow(3) \\
- \; & u_\uparrow(3)u_\downarrow(1)d_\uparrow(2) \\
- \; & u_\uparrow(2)u_\downarrow(3)d_\uparrow(1) \\
- \; & u_\downarrow(1)u_\uparrow(2)d_\uparrow(3) \\
- \; & u_\downarrow(3)u_\uparrow(1)d_\uparrow(2) \\
- \; & u_\downarrow(2)u_\uparrow(3)d_\uparrow(1) \;)
\end{aligned}
\tag{2.11}
$$

(where the antisymmetrisizing sum over the color degrees of freedom has been omitted).

With these definitions the zero momentum propagator (e.g. for a proton with spin up) is given by

$$G_{p\uparrow}(t;\vec{x}_s,t_s) = \sum_{\vec{x}} < \bar{\Psi}_{p\uparrow}(\vec{x},t)\Psi_{p\uparrow}(\vec{x}_s,t_s) > \tag{2.12}$$

and we define the (time dependent) ratio of the g–factors of proton and neutron by

$$\frac{g_p}{g_n}(t) = \frac{\left\{\frac{G_{p\uparrow}-G_{p\downarrow}}{G_{p\uparrow}+G_{p\downarrow}}\right\}_t - \left\{\frac{G_{p\uparrow}-G_{p\downarrow}}{G_{p\uparrow}+G_{p\downarrow}}\right\}_{t-1}}{\text{same but} \quad p \leftrightarrow n} \tag{2.13}$$

In Table I we compare it with the independent quark model result which is $-3/2$ [11]. It shows the (time dependent) ratio for the two types of magnetic fields defined before and for two choices of $x_0$. The data show that:

- for small fields ($eBa^2 \approx 0.01$) the continuum value of $-3/2$ is reproduced very well by all three types.

- for larger fields the '1-ep' type especially for $x_0 = x_s$ is superior.

- for large fields $eBa^2 >\approx 0.05$ lattice discretization and nonlinear effects become important.

As opposed to magnetic moment studies where only energy differences between spin states are important, in the present work, the behavior of the single

| | | magnetic field type | | |
|---|---|---|---|---|
| $eBa^2$ | $t$ | 'fixed' $x_0 = 0$ | '1-ep' $x_0 = 0$ | '1-ep' $x_0 = x_s = L_x/2$ |
| 0.015 | 3 | $-1.47$ | $-1.47$ | $-1.50$ |
| | 4 | $-1.48$ | $-1.48$ | $-1.50$ |
| | 5 | $-1.48$ | $-1.49$ | $-1.50$ |
| | 6 | $-1.49$ | $-1.50$ | $-1.51$ |
| | 7 | $-1.50$ | $-1.52$ | $-1.53$ |
| 0.06 | 3 | $-1.24$ | $-1.24$ | $-1.47$ |
| | 4 | $-1.24$ | $-1.24$ | $-1.46$ |
| | 5 | $-1.29$ | $-1.28$ | $-1.44$ |
| | 6 | $-1.36$ | $-1.35$ | $-1.43$ |
| | 7 | $-1.46$ | $-1.46$ | $-1.42$ |
| 0.12 | 3 | $-1.03$ | $-1.22$ | $-1.42$ |
| | 4 | $-1.04$ | $-1.19$ | $-1.38$ |
| | 5 | $-1.13$ | $-1.23$ | $-1.33$ |
| | 6 | $-1.20$ | $-1.28$ | $-1.30$ |
| | 7 | $-1.27$ | $-1.36$ | $-1.28$ |

Table I: no QCD, $16 \times 16 \times 8 \times 8$, $K = 0.11$: $SU(6)$ (time dependent) ratio of the magnetic moments $g_p/g_n$ $(t)$ for different types of the magnetic field configuration.

'proton' and 'neutron' states in an external magnetic field is of particular interest. The dependence of these states on $B$ in the independent quark model can be computed by applying the Landau formula (1.1) separately to each quark contained in the $SU(6)$ wave function. In particular we find that the proton spin up and the neutron spin down state should not depend in the "flat" limit on $B$ while the states with opposite spin should go up for increasing $B$. In the absence of QCD, the dominant effect is the presence of a contribution $|e_q B|/2m_q$ from the zero-point vibration of each quark.

If we compute these states for different choices of $x_0$ we find a drastic difference between $x_0 = x_s \equiv L_x/2$ and $x_0 = 0$.

Figures 1 and 2 show the effective masses $m_{eff}(t) = \log(G(t+1)/G(t))$ at $t = 5$ (which is representative for all time slices $t$) for the '1-ep' type and $x_0 = 0$ resp. $x_0 = x_s \equiv L_x/2$. While the choice $x_0 = x_s$ reproduces the expected result the choice $x_0 = 0$ shows the protons state becoming heavier than the neutron states for sufficient large $B$. This may mislead to the conclusion that the conjectured effect – the proton getting heavier than the neutron – can be seen

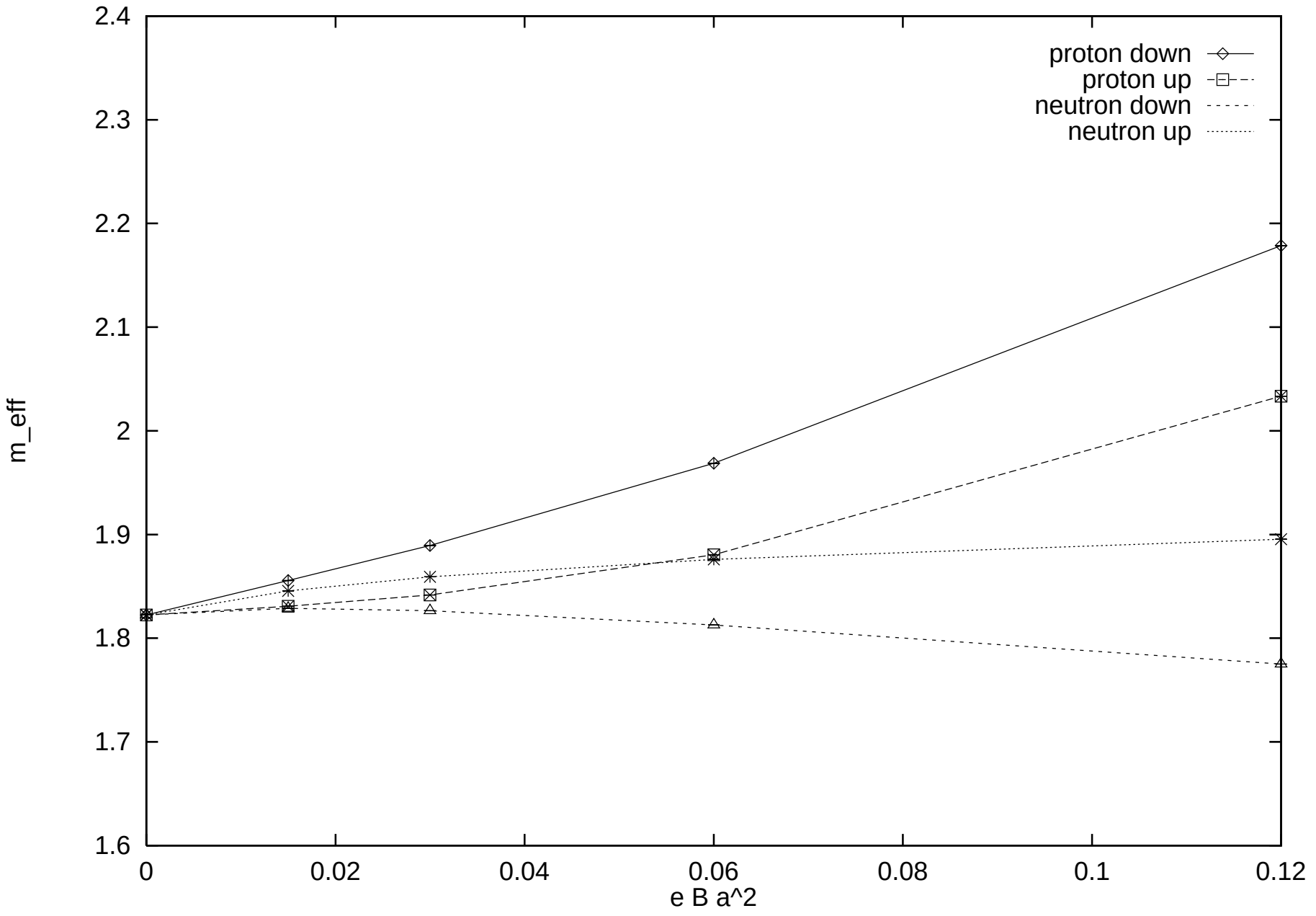

Figure 1: no QCD, $16 \times 16 \times 16 \times 8$: effective masses for proton and neutron in '1-ep' field for $x_0 = 0$. $\diamond = p_\downarrow, \square = p_\uparrow, \triangle = n_\downarrow, \times = n_\uparrow$. see text.

already in the absence of QCD.

However this "effect" is just the result of a gauge dependent definition of the propagators with respect to the electromagnetic gauge transformations. It disappears when the appropriate gauge invariant propagators which take into account the electromagnetic phase-path corrections are used.

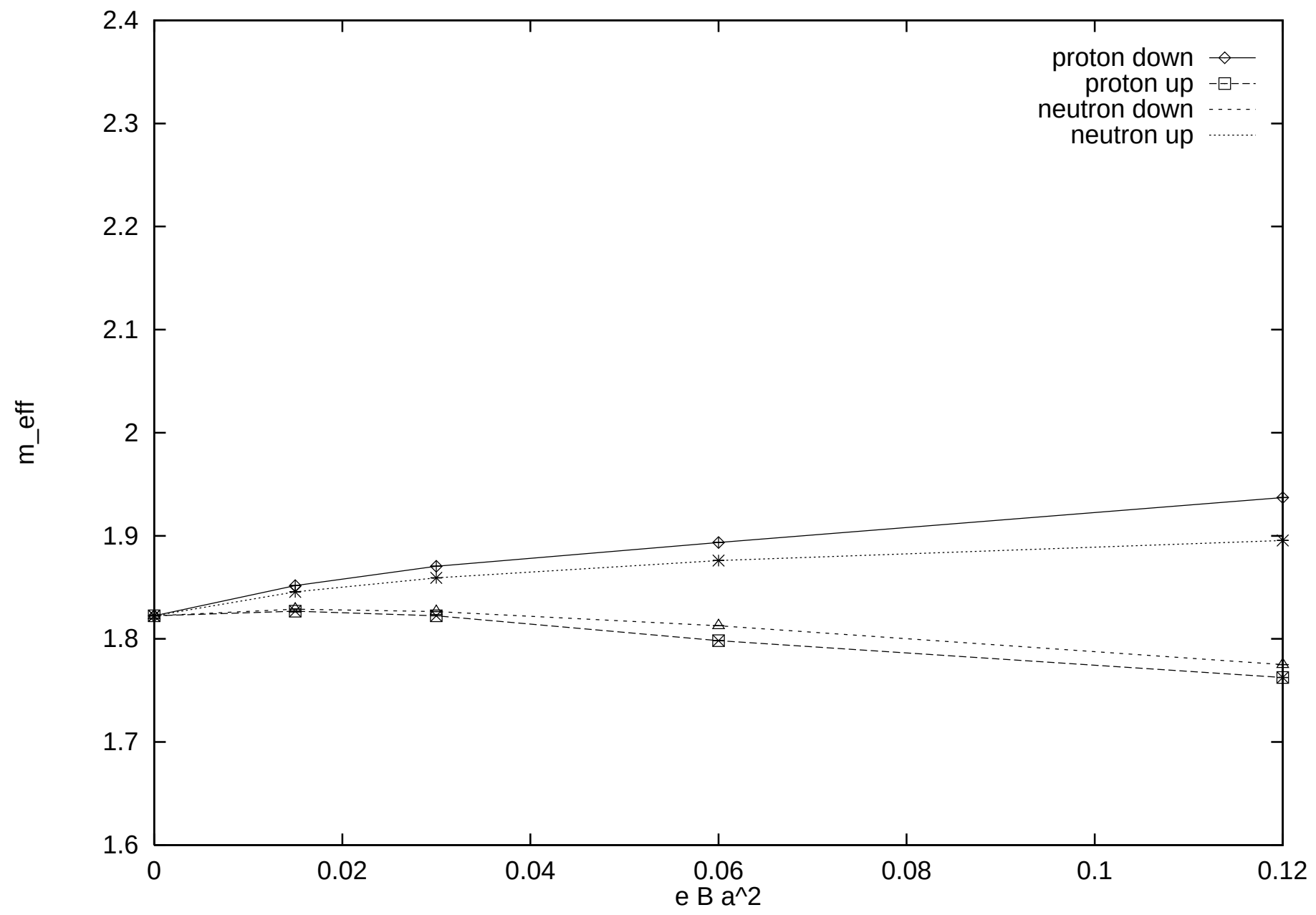

Figure 2: no QCD, $16 \times 16 \times 16 \times 8$: effective masses for proton and neutron in '1-ep' field for $x_0 = x_s$. $\diamond = p_\downarrow, \square = p_\uparrow, \triangle = n_\downarrow, \times = n_\uparrow$. see text.

A hadron propagator invariant under $U(1)$ gauge transformations can be constructed by including a product of $U(1)$ links which lie along a path connecting the source and sink.

For zero momentum it reads

$$G(t; \vec{x}_s, t_s) = \sum_{\vec{x}} < \bar{\Psi}(\vec{x}, t) \hat{U} \Psi(\vec{x}_s, t_s) > \tag{2.14}$$

where $\hat{U}$ denotes the product of $U(1)$ links connecting $\vec{x}, t$ to $\vec{x}_s, t_s$:

$$\hat{U} := U(\vec{x}, t; \vec{x}_n, t_n) \ldots U(\vec{x}_i, t_i; \vec{x}_{i-1}, t_{i-1}) \ldots U(\vec{x}_1, t_1; \vec{x}_s, t_s). \tag{2.15}$$

The exact choice of the path is somewhat arbitrary. However, this is immaterial for ideal QCD simulations where the mass of the lowest state extracted from the propagator (2.14) is unique if the wave function $\Psi(\vec{x}, t)$ has an overlap with the physical state of interest (and if one extracts the mass at sufficiently large $t - t_s$).

The experiments discussed so far have been done with $\hat{U} = 1$. In order to take into account the phase we made experiments with the variants of $\hat{U} \neq 1$. At the beginning we still discuss the case where QCD is switched off and we can compare with the continuum independent quark model predictions. As an example we choose for the path the shortest path between $(\vec{x}_s, t_s)$ and $(\vec{x}, t)$. The nontrivial contribution in (2.15) is then coming from

$$U_y(\vec{x}, t) = \exp(ieBa^2(x - x_0)) \tag{2.16}$$

This corresponds to both types of the magnetic field '1-ep' (2.6) and 'fixed' (2.8) but with the $x$–links at the boundary omitted. In this way we get

$$\hat{U} = \exp iBa^2(x - x_0)(y - y_s) \sum_q e_q \tag{2.17}$$

where the sum runs over all charges $e_q$ of the quarks contained in the hadron of interest.

We find that this choice

- does not change (almost) the behavior for $x_0 = x_s$.
- corrects the wrong behavior for $x_0 = 0$ such that it becomes identical with the case $x_0 = x_s$

The results discussed so far imply that for the 'flat' case the lattice represents the continuum if we use the '1-ep' type and $x_0 = x_s$ or include the path in the propagator as described before. Note that due to the Polyakov loop (finite size) effects, the 2 choices could have given in principle different results.

This choice has been used for all the results presented in the sequel.

## 2.2 Wave functions for proton and neutron

In order to extract hadron masses from the time dependence of the correlation function the wave function for the particle of interest should have maximal overlap with the physical state. One way to achieve a better overlap is certainly to use non pointlike sources [12]. Due to the interaction of the magnetic field type with the position of the source we have considered the simplest case of pointlike sources.

We have found that the overlap is much more sensitive to the choice of the wave function than it is for 'non–magnetic' mass simulations.

In order to find a wave function which is suitable for the present work we were again guided by the case where QCD is switched off.

It is well known that the independent quark model predicts a ratio for the g–factors of proton and neutron, $g_p/g_n = -3/2$ [11], which is remarkably close to the experimental value of $-1.47$. The value of $-3/2$ can be computed from the magnetic moments of the 'proton' and 'neutron' states given by the $SU(6)$ wave function.

In order to estimate the reliability of our numerical results and their sensitivity to the changes in the wave function we compared the $SU(6)$ results with the ones obtained with the $(u\gamma_5 Cu)d$ wave function.

In the absence of electric charge effects, the two wave functions are equivalent and $(u\gamma_5 Cu)d$ is the usually used wave function. In the presence of electric charge effects the $SU(6)$ function is the correct one, while $(u\gamma_5 Cu)d$ gives the incorrect answer $-5/4$.

We used these differences for checking the consistency and reliability of our numerical procedures. We have computed the (time dependent) ratio of the g-factors for proton and neutron (2.13) for the two choices of the wave function: $(uC\gamma_5 u)d$ and $SU(6)$. Table II shows the results obtained on a $16 \times 16 \times 8 \times 8$ lattice from 20 quenched configurations at at $\beta = 6.0$, the '1-ep' type magnetic field with $x_0 = x_s$ and $K = 0.1475$. For small magnetic fields $eBa^2 \leq 0.02$ the $SU(6)$ data are close to the experimental value of $-1.47$. However the data for the $(uC\gamma_5 u)d$ wave function are close to the $-5/4$ of the flat case. They increase for large time–slices and thus indicate a mixing with higher states. This implies of course that the $(uC\gamma_5 u)d$ wave function has a poorer overlap with the physical state than the $SU(6)$ wave function. From the table we also can see that for larger fields $eBa^2 > O(0.01)$ the ratio decreases with increasing magnetic field. This has already been seen in the flat case (Table I).

| $eBa^2$ | $t$ | $g_p/g_n\ (t)$ | |
|---|---|---|---|
| | | $(uC\gamma_5 u)d$ | $SU(6)$ |
| 0.01 | 3 | $-1.22$ | $-1.48$ |
| | 4 | $-1.23$ | $-1.52$ |
| | 5 | $-1.26$ | $-1.52$ |
| | 6 | $-1.26$ | $-1.50$ |
| | 7 | $-1.29$ | $-1.54$ |
| 0.02 | 3 | $-1.21$ | $-1.48$ |
| | 4 | $-1.23$ | $-1.52$ |
| | 5 | $-1.27$ | $-1.52$ |
| | 6 | $-1.26$ | $-1.50$ |
| | 7 | $-1.27$ | $-1.53$ |
| 0.08 | 3 | $-1.18$ | $-1.45$ |
| | 4 | $-1.18$ | $-1.48$ |
| | 5 | $-1.24$ | $-1.48$ |
| | 6 | $-1.18$ | $-1.45$ |
| | 7 | $-1.07$ | $-1.38$ |
| 0.15 | 3 | $-1.14$ | $-1.41$ |
| | 4 | $-1.10$ | $-1.38$ |
| | 5 | $-1.10$ | $-1.33$ |
| | 6 | $-0.99$ | $-1.28$ |
| | 7 | $-0.79$ | $-1.10$ |

Table II: QCD, $16 \times 16 \times 8 \times 8$, $K = 0.1475$, '1-ep', $x_0 = x_s$: (time dependent) ratio of the magnetic moments $g_p/g_n\ (t)$ for the $(uC\gamma_5 u)d$ and the $SU(6)$ wave function.

# 3 Results

In this section we present the data obtained from the numerical simulations. In particular we have computed the slopes

$$slope(B) = (m(B) - m(0))\frac{2m(B)}{eB}, \tag{3.1}$$

the g–factors for proton and neutron and the ratio of their magnetic moments.

Table $III$ shows the numerical results in comparison with the experimental data. The upper two parts of the table contain the data for an order of magnitude for the magnetic field, where – according to the mechanism described in the introduction – the proton should become heavier than the neutron. The lower third of the table contains the data for a larger magnetic field, $eB = 0.06$ resp. $eB = 0.08$ (These data stem from 20 configurations). The results for the different values of $B$ are the same within the error bars. In the previous section we have already seen that for larger magnetic fields our lattice does not reproduce the precise continuum results even in the absence of QCD.

The data stem from a $16 \times 16 \times 8 \times 8$ lattice at $K = 0.1475$ and $0.1525$ and from a $24 \times 16 \times 16 \times 12$ lattice at $K = 0.145$ and $0.15$. The data differ for these two lattice sizes. In particular the spin up slopes differ by amounts which can hardly be explained by the error bars. If we extrapolate the data to infinite volume by assuming a $a = a_0 + 1/Va_1$ behavior we get results which are remarkably close to the experimental data (Table III columns 2 and 3). Moreover this extrapolation brings agreement with the proton-decay scenario of [1]. To see that, compare in Table III, column 3 the infinite volume extrapolation of energy/B slope of the lowest nucleon states: while for the neutron the energy decreases with slope $-2.1$ ( neutron $\downarrow$ state in line 1) for the proton energy decreases only with slope $-1.9$ (proton $\uparrow$ state in line 5). Still we refrain from considering these results as definite and reliable confirmation of the theoretical prediction of the proton decay [1]. For such a confirmation larger lattices with better statistics will be necessary.

The statistical errors have been computed by dividing the configurations into bins and comparing the results for each of them. The systematic errors are hard to estimate. The most significant one certainly stems from the use of pointlike sources. They make it difficult to avoid the systematic overestimation of the masses caused by higher excitations of the spectrum. We have estimated this effect by extracting the masses from the correlation function for different windows of time–slices.

The ratio of the g–factors does not suffer from the former effects, since they consist of ratios of masses. As already presented in Table II the ratios (2.13) are very stable for a quite large numbers of time slices.

| | phenom./ exper. | $V=\infty$ | $K=0.145$ $eB=0.03$ $24\times16\times16\times12$ | $K=0.1475$ eB=0.02 $16\times16\times8\times8$ |
|---|---|---|---|---|
| neutron ↓ slope | $-1.9$ | $-2.1$ | $-2.1(4)$ | $-2.2(4)$ |
| neutron ↑ slope | $+1.9$ | $+1.3$ | $+1.9(3)$ | $+3.2(2)$ |
| neutron g–factor $g_n$ | $-1.9$ | $-1.7$ | $-2.0(4)$ | $-2.6(4)$ |
| proton ↓ slope | $+3.8$ | $+3.5$ | $+3.4(4)$ | $+3.3(4)$ |
| proton ↑ slope | $-1.8$ | $-1.9$ | $-2.9(4)$ | $-4.8(2)$ |
| proton g-factor $g_p$ | $+2.8$ | $+2.7$ | $+3.1(4)$ | $+4.1(5)$ |
| $g_p/g_n$ | $-1.47$ | $-1.50$ | $-1.51(3)$ | $-1.53(4)$ |

| | phenom./ exper. | $V=\infty$ | $K=0.15$ $eB=0.03$ $24\times16\times16\times12$ | $K=0.1525$ eB=0.02 $16\times16\times8\times8$ |
|---|---|---|---|---|
| neutron ↓ slope | $-1.9$ | $-1.8$ | $-2.4(4)$ | $-3.5(8)$ |
| neutron ↑ slope | $+1.9$ | $+1.5$ | $+2.0(4)$ | $+3.0(7)$ |
| neutron g–factor $g_n$ | $-1.9$ | $-1.7$ | $-2.2(4)$ | $-3.3(8)$ |
| proton ↓ slope | $+3.8$ | $+3.5$ | $+3.8(4)$ | $+4.5(8)$ |
| proton ↑ slope | $-1.8$ | $-1.1$ | $-3.2(4)$ | $-7.4(8)$ |
| proton g-factor $g_p$ | $+2.8$ | $+2.3$ | $+3.5(4)$ | $+6.0(8)$ |
| $g_p/g_n$ | $-1.47$ | $-1.44$ | $-1.53(8)$ | $-1.7(2)$ |

| | phenom./ exper. | $V=\infty$ | $K=0.145$ $eB=0.06$ $24\times16\times16\times12$ | $K=0.1475$ eB=0.08 $16\times16\times8\times8$ |
|---|---|---|---|---|
| neutron ↓ slope | $-1.9$ | $-2.2$ | $-2.5(5)$ | $-3.1(6)$ |
| neutron ↑ slope | $+1.9$ | $+1.3$ | $+1.5(3)$ | $+2.0(3)$ |
| neutron g–factor $g_n$ | $-1.9$ | $-1.7$ | $-2.0(5)$ | $-2.6(6)$ |
| proton ↓ slope | 3.8 | $+3.3$ | $+3.2(4)$ | $+3.0(6)$ |
| proton ↑ slope | $-1.8$ | $-1.7$ | $-2.8(4)$ | $-4.9(3)$ |
| proton g-factor $g_p$ | 2.8 | $+2.5$ | $+3.0(4)$ | $+4.0(6)$ |
| $g_p/g_n$ | $-1.47$ | $-1.48$ | $-1.50(4)$ | $-1.54(5)$ |

Table III: slopes of proton and neutron states, g–factors and their ratios.

Figure 3 shows the proton and hadron masses as a function of the magnetic field for the $24 \times 16 \times 16 \times 12$ lattice and $K = 0.145$.

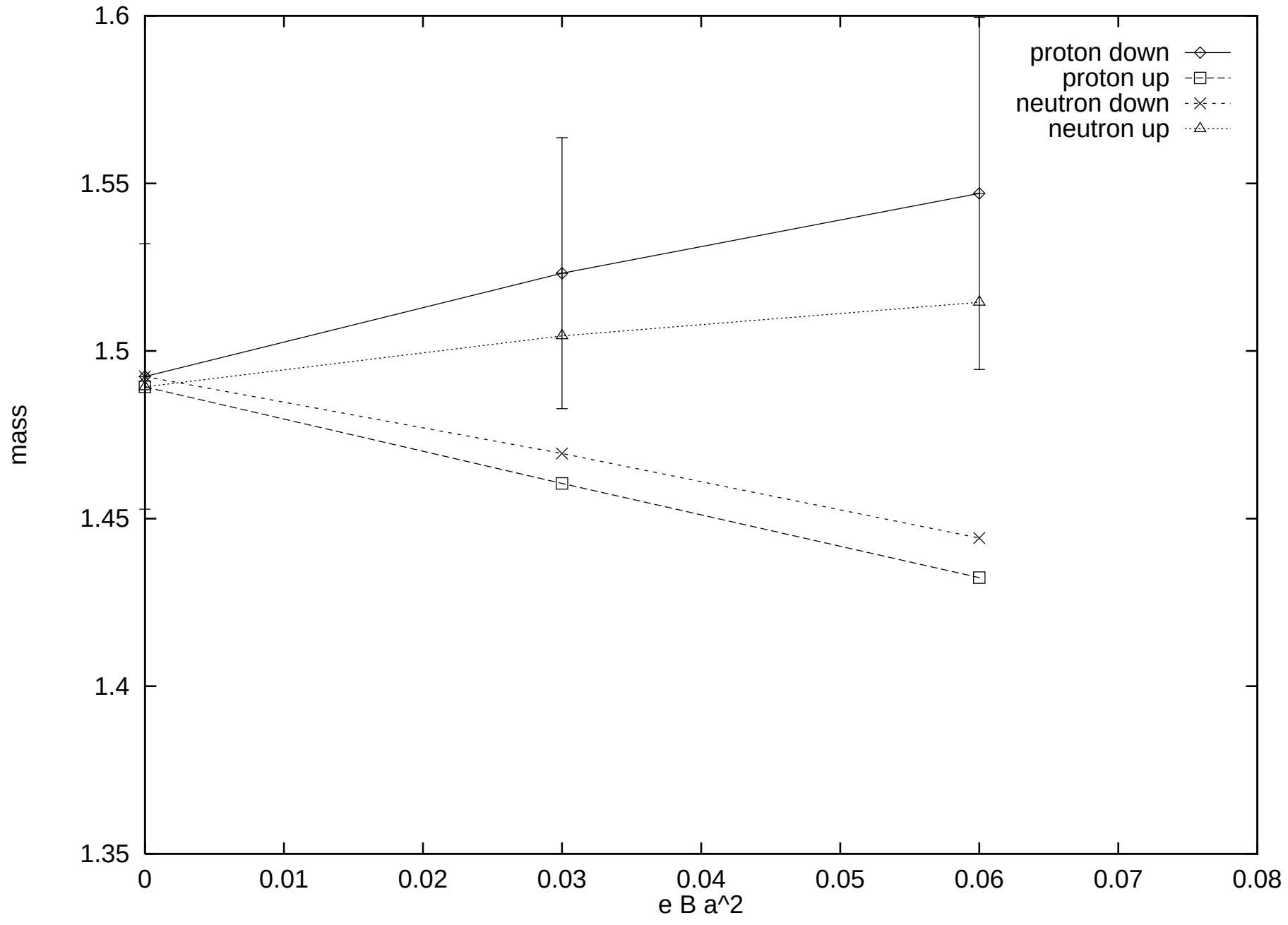

Figure 3: QCD, $24\times16\times16\times12$: masses for proton and neutron for $K = 0.145$. For better readability error bars are shown only for one state.

# 4 Conclusions and Outlook

In the present work we showed which ingredients are important in order to study the variation of hadron masses in an external magnetic field.

In particular we pointed out the factors influencing the lattice results:

- the choice of the (SU(6)) wave function.
- the location and the boundary discontinuity of the magnetic field.
- the path-phase correction which makes the charged-particles propagators QED-gauge-invariant.
- the location of the source

We computed the slopes (mass/magnetic field) for the lowest spin up and down states of proton and neutron and compared them with the phenomenological approach of [1].

We worked at very large magnetic fields which are expected to drive the proton to a larger mass than the neutron. We found the remarkable result that –within the available numerical precision– the assumptions leading to (1.1) and (1.2) hold for these very large fields. However in order to decide whether the proton becomes heavier than the neutron more precision is necessary.

Further investigations thus should include:
- a carefull study of the finite size effects
- better statistics
- smeared sources adapted to the charged particles problem
- extrapolation to the chiral limit.

## Acknowledgements

We thank P.G. Lauwers and V. Rittenberg for useful discussions.

T.W. would like to thank for the warm hospitality during his visits at the Hebrew University of Jerusalem and the University of Uppsala. Financial support by the *Deutsche Forschungsgemeinschaft* and EEC SCIENCE contract no SC1 CT910674 is gratefully acknowledged.

We also acknowledge support from the Germany Israel Foundation (GIF) and from the Fundation for Fundamental Research of the Israeli Academy.

The numerical computations have been performed at the *Nationellt Superdatorcentrum*, Linköping, Sweden and at the *Höchstleistungsrechenzentrum* (HLRZ), Jülich, Germany.